\documentclass[aps,onecolumn,showpacs,nofootinbib,showkeys]{revtex4-2}
\usepackage[margin=3.0cm]{geometry}

\RequirePackage[T1]{fontenc}

\usepackage{epsfig,graphicx,amsmath,amssymb,bm}
\RequirePackage{mathptmx}      
\usepackage{mathrsfs}
\usepackage{amssymb}
\RequirePackage{color}
\usepackage{changes}
\usepackage{slashed}
\usepackage{multirow}
\usepackage{scalerel}
\usepackage{tikz-feynman}
\usepackage{textcomp}
\usepackage{subcaption}
\usepackage[english]{babel}
\usepackage{float}
\RequirePackage{hyperref}
\hypersetup{
    linktocpage,
    colorlinks,
    citecolor=blue,
    filecolor=black,
    linkcolor=blue,
    urlcolor=blue,
}
\usepackage{nccmath}

\begin{document}

\title{Description of the decay $\phi \to \pi^+\pi^-$ in the NJL chiral quark model}


\author{M.K. Volkov$^{1}$}\email{volkov@theor.jinr.ru}
\author{A.A. Pivovarov$^{1}$}\email{pivovarov@theor.jinr.ru}
\author{K. Nurlan$^{1,2}$}\email{nurlan@theor.jinr.ru}

\affiliation{$^1$ Bogoliubov Laboratory of Theoretical Physics, JINR, 
                 141980 Dubna, Moscow region, Russia \\
                $^2$ The Institute of Nuclear Physics, Almaty, 050032, Kazakhstan}   


\begin{abstract}
The doubly suppressed decay $\phi \to \pi^+\pi^-$ is described in the chiral NJL quark model. This decay occurs in violation of the OZI rule and isotopic invariance. Two possible channels are considered. The first channel involves $\phi-\omega$ mixing and the subsequent decay of $\omega \to \pi^+\pi^-$. In the second channel, the decay occurs via an intermediate $\rho$ meson.
\end{abstract}

\pacs{}

\maketitle


\section{\label{Intro}Introduction}
Vector meson decays provide a good laboratory for testing theoretical models of the strong interaction and studying the structure of hadrons. Among vector mesons, special attention is paid to the $\phi$-meson consisting of $s\bar{s}$ with quantum numbers $J^{PC}=1^{--}$, mass $M_\phi=1019.460 \pm 0.016$~MeV and a relatively narrow width $\Gamma_\phi = 4.249 \pm 0.013$~MeV \cite{ParticleDataGroup:2026mpi}. The accumulated data with large statistics at electron-positron colliders and $\phi$-factories make this resonance unique for precision measurements.

The standard representation of the quark-antiquark structure of the $\phi$ meson is in good agreement with the quark model and the Okubo - Zweig  Iizuka (OZI) rule, where the dominant channels are decays into a pair of kaons $K^+K^-$ and $K_S^0 K_L^0$ \cite{ParticleDataGroup:2026mpi}. At the same time, decays that occur in violation of the OZI rule are of great interest. Examples of such decays are $\phi \to \pi\gamma, \rho\pi, 3\pi$, which can be described by the transition $\phi \to \omega$ \cite{Klingl:1996by,Benayoun:2007cu,Danilkin:2014cra,Volkov:2020jor}.

It is worth noting the strongly suppressed $\phi \rightarrow \pi^{+}\pi^{-}$ process, where, in addition to the violation of the OZI rule, an additional constraint arises due to the violation of isotopic invariance. In the $\phi \rightarrow \pi^{+}\pi^{-}$ decay, as in the $\omega \rightarrow \pi^{+}\pi^{-}$ case, the difference in the masses of the $u$ and $d$ quarks plays an important role. This mass difference appears directly in the quark loop integrals in the $\omega \rightarrow \pi^{+}\pi^{-}$ decay, and in the $\phi \rightarrow \rho \rightarrow \pi^{+}\pi^{-}$ process, it appears in the masses of charged and neutral kaons in the meson loop of the $\phi$--$\rho$ transition. It is important to note that the difference in the masses of the $K^{\pm}$ and $K^{0}$ kaons is mainly determined by the difference in the masses of the $u$ and $d$ quarks~\cite{Volkov:1986zb}.

The experimental situation with the decay $\phi \rightarrow \pi^{+}\pi^{-}$ remains uncertain due to insufficient measurement accuracy. Despite the accumulated data from the CMD-3 and SND detectors at VEPP-2000, the statistics of this process remain low compared to the dominant decay modes $(\phi \rightarrow KK)$~\cite{Achasov:2000pu,CMD-3:2023alj}. The decay branching fraction $Br(\phi \rightarrow \pi \pi) = (9.5 \pm 1.9) \times 10^{-5}$ given in the PDG is based on the statistics accumulated in the $\phi$ resonance region at electron-positron colliders. This situation stimulates theoretical studies, in particular, the testing of various mechanisms of $\phi$--$\omega$ and $\phi$--$\rho$ mixing.

The present paper is devoted to a theoretical study of the $\phi \rightarrow \pi^{+}\pi^{-}$ decay in the Nambu--Jona-Lasinio model~\cite{Nambu:1961tp,Eguchi:1976iz,Ebert:1982pk,Volkov:1984kq,Volkov:1986zb,Ebert:1985kz,Vogl:1991qt,Klevansky:1992qe,Volkov:1993jw,Ebert:1994mf,Hatsuda:1994pi,Buballa:2003qv,Volkov:2005kw}. As a mechanism for the decay, we consider the mixing of $\phi$--$\omega$ states with a subsequent decay into two pions, where the $\omega \rightarrow \pi^{+}\pi^{-}$ decay includes contributions from quark loops, the electromagnetic transition, and meson loops. In addition, we consider the direct $\phi$--$\rho$ transition, which occurs due to the difference in meson loops with neutral and charged kaons.

\section{Lagrangian of the NJL model}
In this work, we employ the effective Lagrangian of the NJL model with $U(3) \times U(3)$ symmetry, which describes the interactions between quarks and mesons and is derived via the bosonization of a local four-quark chiral Lagrangian motivated by QCD. The derivation of a quark-meson Lagrangian based on the effective four-quark Lagrangian of the NJL model is described in \cite{Ebert:1982pk,Volkov:1984kq,Volkov:1986zb,Volkov:2005kw}. This model, based on the spontaneous breaking of chiral symmetry, has proven successful in calculating a large number of processes in low-energy meson physics. The effective quark-meson Lagrangian of the model, containing the vertices necessary for calculating the process under consideration, takes the form:
\begin{eqnarray}
	\label{Lagrangian}
		\Delta L_{int} & = &
		\bar{q} \biggl[ i g_{\pi} \gamma^{5} \sum_{j = \pm} \lambda_{j}^{\pi} {\pi}^{j}
		+\frac{g_{\rho}}{2} \gamma^{\mu} \lambda^{\rho} \rho^{0}_{\mu} + \frac{g_{\omega}}{2} \gamma^{\mu} \lambda^{\omega} \omega_{\mu} + \frac{g_{\phi}}{2} \gamma^\mu\lambda^\phi \phi_\mu \nonumber\\
        && + ig_K\gamma^5\sum_{j=\pm}\lambda_j^K K^j + ig_K\gamma^5\lambda_0^K K^0 + ig_K\gamma^5\lambda_0^{\bar{K}} \bar{K}^0
		\biggl]q,
\end{eqnarray}
where $q$ and $\bar{q}$ are the triplets of the u, d and s quarks, $\lambda$ are the linear combinations of the Gell-Mann matrices.

The coupling constants for the interaction of mesons with quarks are obtained as a result of redefining the meson fields and take the following form~\cite{Volkov:1986zb}:
\begin{eqnarray}
 g_{\pi} = \left(\frac{4}{Z_{\pi}}I_{20}\right)^{-1/2}, \quad
 g_{\rho} = g_{\omega} = \left(\frac{2}{3}I_{20}\right)^{-1/2}, \quad
 g_{\phi} = \left(\frac{2}{3}I_{02}\right)^{-1/2}, \quad
 g_K = \left(\frac{4}{Z_{K}}I_{11}\right)^{-1/2}.
\end{eqnarray}
where $Z_{\pi}$, $Z_K$ are additional renormalization constants that arise when transitions between an axial-vector meson and a pseudoscalar meson are taken into account:
\begin{eqnarray}
    &Z_{\pi} = \left(1 - 6\frac{m_{u}^{2}}{M^{2}_{a_{1}}}\right)^{-1}, \quad Z_{K} = \left(1 - \frac{3}{2}\frac{\left(m_{u} + m_s\right)^{2}}{M^{2}_{K_{1A}}}\right)^{-1},& \nonumber\\
    &M^{2}_{K_{1A}} = \left(\frac{\sin^{2}{\alpha}}{M^{2}_{K_{1}(1270)}} + \frac{\cos^{2}{\alpha}}{M^{2}_{K_{1}(1400)}}\right)^{-1},&
\end{eqnarray}
where $m_u = 270$~MeV and $m_s = 420$~MeV are the constituent mass of the $u$ and $s$ quarks, $M_{a_{1}}$ is the mass of the $a_1$ meson, $M_{K_{1}(1270)}$ and $M_{K_{1}(1400)}$ are the masses of the $K_1$ mesons, and the mixing angle $\alpha = 57^{\circ}$~\cite{Volkov:2019awd}.

The integrals that appear in quark loops during the renormalization of the Lagrangian take the form
\begin{eqnarray}
\label{integral_1}
	I_{n_{1}n_{2}} =
	-i\frac{N_{c}}{(2\pi)^{4}}\int\frac{\Theta(\Lambda_{4}^{2} + k^2_{\perp})}{(m_{u}^{2} - k^2)^{n_{1}}(m_{s}^{2} - k^2)^{n_{2}}}
	\mathrm{d}^{4}k,
\end{eqnarray}
where $\Lambda_4=1265$~MeV is the cut-off parameter~\cite{Volkov:1986zb}.

\section{Amplitude of the process $\phi \to \pi^+ \pi^-$}
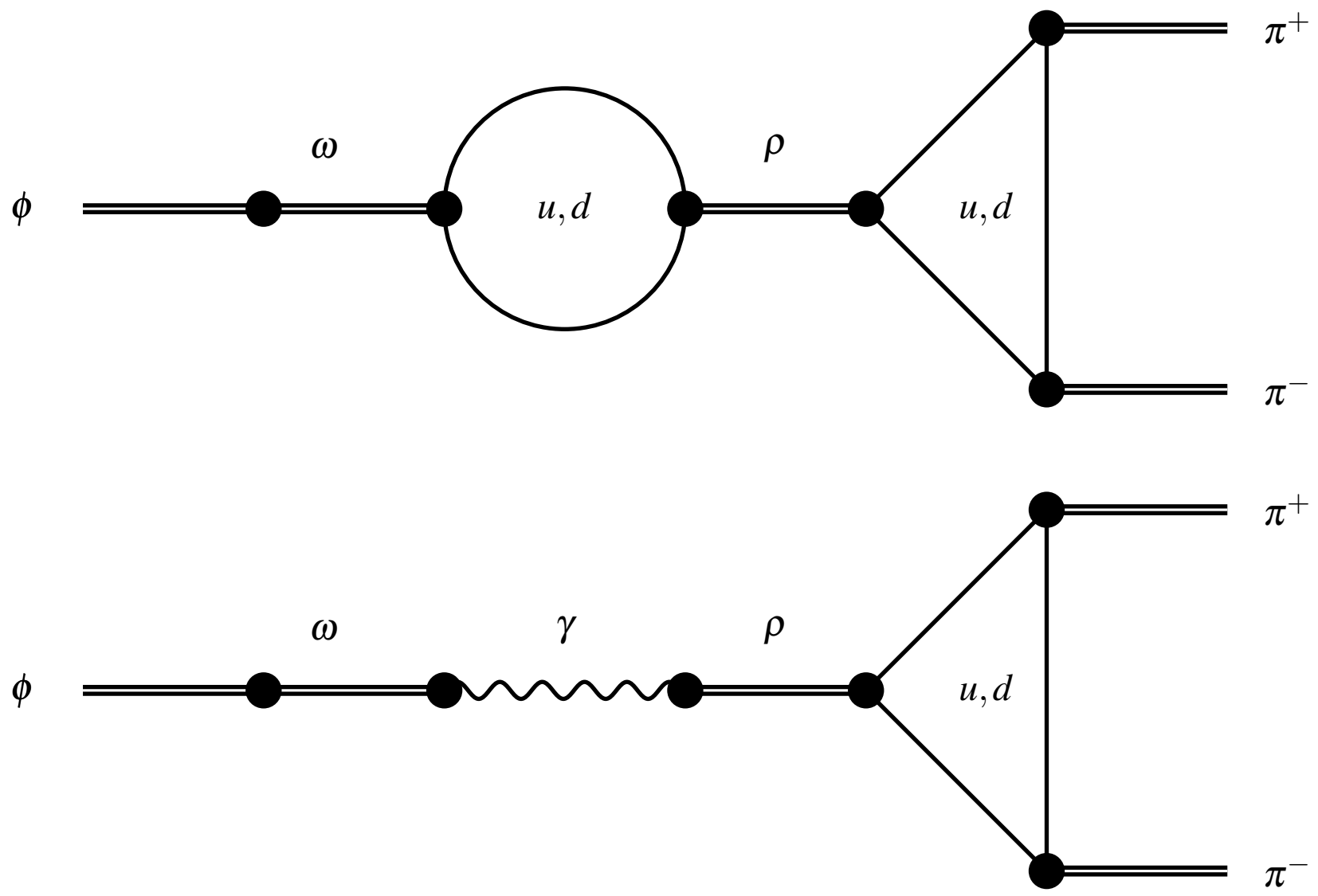
\begin{figure*}[t]
\begin{tikzpicture}
  \draw[draw={rgb,255:red,0;green,0;blue,0}, line width=1.00pt, double] (1.00, 10.00) -- (2.50, 10.00);
  \draw[draw={rgb,255:red,0;green,0;blue,0}, line width=1.00pt, double] (2.50, 10.00) -- (4.00, 10.00);
  \draw[draw={rgb,255:red,0;green,0;blue,0}, line width=1.00pt, double] (6.00, 10.00) -- (7.50, 10.00);
  \draw[draw={rgb,255:red,0;green,0;blue,0}, line width=1.00pt] (7.50, 10.00) -- (9.00, 11.50);
  \draw[draw={rgb,255:red,0;green,0;blue,0}, line width=1.00pt] (7.50, 10.00) -- (9.00, 8.50);
  \draw[draw={rgb,255:red,0;green,0;blue,0}, line width=1.00pt] (9.00, 11.50) -- (9.00, 8.50);
  \draw[draw={rgb,255:red,0;green,0;blue,0}, line width=1.00pt, double] (9.00, 11.50) -- (10.50, 11.50);
  \draw[draw={rgb,255:red,0;green,0;blue,0}, line width=1.00pt, double] (9.00, 8.50) -- (10.50, 8.50);
  \draw[fill=white!20, draw={rgb,255:red,0;green,0;blue,0}, line width=1.00pt] (5.00, 10.00) ellipse (1.00 and 1.00);
  \draw[draw={rgb,255:red,0;green,0;blue,0}, line width=1.00pt, double] (1.00,   6.00) -- (2.50, 6.00);
  \draw[draw={rgb,255:red,0;green,0;blue,0}, line width=1.00pt, double] (2.50, 6.00) -- (4.00, 6.00);
  \draw[draw={rgb,255:red,0;green,0;blue,0}, line width=1.00pt, double] (6.00, 6.00) -- (7.50, 6.00);
  \draw[draw={rgb,255:red,0;green,0;blue,0}, line width=1.00pt] (7.50, 6.00) -- (9.00, 7.50);
  \draw[draw={rgb,255:red,0;green,0;blue,0}, line width=1.00pt] (7.50, 6.00) -- (9.00, 4.50);
  \draw[draw={rgb,255:red,0;green,0;blue,0}, line width=1.00pt] (9.00, 7.50) -- (9.00, 4.50);
  \draw[draw={rgb,255:red,0;green,0;blue,0}, line width=1.00pt, double] (9.00, 7.50) -- (10.50, 7.50);
  \draw[draw={rgb,255:red,0;green,0;blue,0}, line width=1.00pt, double] (9.00, 4.50) -- (10.50, 4.50);
  \draw[draw={rgb,255:red,0;green,0;blue,0}, line width=1.00pt, decorate, decoration={snake, segment length=10pt, amplitude=2pt}] (4.00, 6.00) -- (6.00, 6.00); 
  \fill (2.50, 10.00) circle (0.15);
  \fill (4.00, 10.00) circle (0.15);
  \fill (6.00, 10.00) circle (0.15);
  \fill (7.50, 10.00) circle (0.15);
  \fill (9.00, 11.50) circle (0.15);
  \fill (9.00, 8.50) circle (0.15);
  \node at (0.50, 10.00) {$\phi$};
  \node at (3.00, 10.50) {$\omega$};
  \node at (6.75, 10.50) {$\rho$};
  \node at (11.00, 11.50) {$\pi^+$};
  \node at (11.00, 8.50) {$\pi^-$};
  \node at (0.50, 10.00) {$\phi$};
  \node at (5.00, 10.00) {$u,d$};
  \node at (8.50, 10.00) {$u,d$};
  \fill (2.50, 6.00) circle (0.15);
  \fill (4.00, 6.00) circle (0.15);
  \fill (6.00, 6.00) circle (0.15);
  \fill (7.50, 6.00) circle (0.15);
  \fill (9.00, 7.50) circle (0.15);
  \fill (9.00, 4.50) circle (0.15);
  \node at (0.50, 6.00) {$\phi$};
  \node at (3.00, 6.50) {$\omega$};
  \node at (6.75, 6.50) {$\rho$};
  \node at (11.00, 7.50) {$\pi^+$};
  \node at (11.00, 4.50) {$\pi^-$};
  \node at (0.50, 6.00) {$\phi$};
  \node at (5.00, 6.50) {$\gamma$};
  \node at (8.50, 6.00) {$u,d$};
\end{tikzpicture}
  \caption{Diagrams describing the decay of $\phi\to\pi^+\pi^-$ via $\phi-\omega$ mixing}
  \label{omegapipi}
\end{figure*}%

The processes $\phi\to\pi\gamma$ and $\phi\to3\pi$ that violate the OZI rule violation are studied in the NJL model \cite{Volkov:2020jor}. These decays are described by mixing the isoscalar $\phi$ and $\omega$ states.
The mixing angle $\alpha=3.1^\circ$ was fixed based on the experimental decay width $\Gamma(\phi\to\pi\gamma)=5.65\pm0.21$ keV.
This value was used in the description of the decay $\phi\to3\pi$, where good agreement was obtained for the decay width $\Gamma(\phi\to3\pi)=0.67$ MeV with the experimental data $\Gamma(\phi\to3\pi)_{exp.}=0.68\pm0.03$ MeV.
In addition, the angle value is in agreement with the results of \cite{Klingl:1996by,Benayoun:2007cu,Ambrosino:2009sc}. In this work, we use this mixing to describe the decay $\phi \to \omega\to \pi^+ \pi^-$. It should be noted that the decay of $\omega\to \pi^+ \pi^-$ is mainly possible due to the difference in the masses of $u$- and $d$-quarks \cite{Volkov:1986zb}.

Diagrams describing this channel for the decay $\phi\to\pi^+\pi^-$ are presented in Fig. \ref{omegapipi}. The amplitude corresponding to these diagrams takes the form
\begin{eqnarray}
\label{phiomega}
    \mathcal{M}_{\phi\omega} = - \left[ \frac{g^3_\rho}{3} 
    \left[I_{20}(u) - I_{20}(d)\right]  \,
    - \frac{4 \pi \alpha_{em}}{3 g_\rho} 
    \right] M^2_\phi \sin{\alpha}
    BW_{\rho}(p)
    e_{\mu}^*(p) (p_{\pi^+}-p_{\pi^-})^\mu,
\end{eqnarray}
where $p_{\pi^+}$ and $p_{\pi^-}$ are the pion momenta in the final state, $e_{\mu}^*(p)$ is the $\phi$-meson polarization vector, $I_{20}(u)$ and $I_{20}(d)$ are the quark loop integrals defined in (\ref{integral_1}), and $m_d - m_u = 4$ MeV \cite{Volkov:1986zb,Volkov:2005kw}.

Intermediate mesons are described by the Breit-Wigner propagator
\begin{eqnarray}
    BW_{\rho}(q) = \frac{1}{M_{\rho} - q^2 - i\sqrt{q^2}\Gamma_{\rho}}.
\end{eqnarray}

As a result, the contribution of the $\phi-\omega$ mixing channel to the decay branching fractions $\phi\to \pi^+ \pi^-$ is
\begin{eqnarray}
    Br(\phi\to \pi^+ \pi^-)_{\phi\omega} = 2.8 \times 10^{-5}.
\end{eqnarray}

\begin{figure*}[t]
\begin{tikzpicture}
  \draw[draw={rgb,255:red,0;green,0;blue,0}, line width=1.00pt, double] (2.50, 6.00) -- (4.00, 6.00);
  \draw[draw={rgb,255:red,0;green,0;blue,0}, line width=1.00pt, double] (6.00, 6.00) -- (7.50, 6.00);
  \draw[draw={rgb,255:red,0;green,0;blue,0}, line width=1.00pt] (7.50, 6.00) -- (9.00, 7.50);
  \draw[draw={rgb,255:red,0;green,0;blue,0}, line width=1.00pt] (7.50, 6.00) -- (9.00, 4.50);
  \draw[draw={rgb,255:red,0;green,0;blue,0}, line width=1.00pt] (9.00, 7.50) -- (9.00, 4.50);
  \draw[draw={rgb,255:red,0;green,0;blue,0}, line width=1.00pt, double] (9.00, 7.50) -- (10.50, 7.50);
  \draw[draw={rgb,255:red,0;green,0;blue,0}, line width=1.00pt, double] (9.00, 4.50) -- (10.50, 4.50);
  \draw[draw={rgb,255:red,0;green,0;blue,0}, line width=1.00pt, decorate, decoration={snake, segment length=10pt, amplitude=2pt}] (4.00, 6.00) -- (6.00, 6.00); 
  \fill (4.00, 6.00) circle (0.15);
  \fill (6.00, 6.00) circle (0.15);
  \fill (7.50, 6.00) circle (0.15);
  \fill (9.00, 7.50) circle (0.15);
  \fill (9.00, 4.50) circle (0.15);
  \node at (2.00, 6.00) {$\phi$};
  \node at (6.75, 6.50) {$\rho$};
  \node at (11.00, 7.50) {$\pi^+$};
  \node at (11.00, 4.50) {$\pi^-$};
  \node at (5.00, 6.50) {$\gamma$};
  \node at (8.50, 6.00) {$u,d$};
\end{tikzpicture}
  \caption{Diagram describing the decay $\phi\to\pi^+\pi^-$ by electromagnetic $\phi-\rho$ mixing with photon exchange}
  \label{phigamma}
\end{figure*}
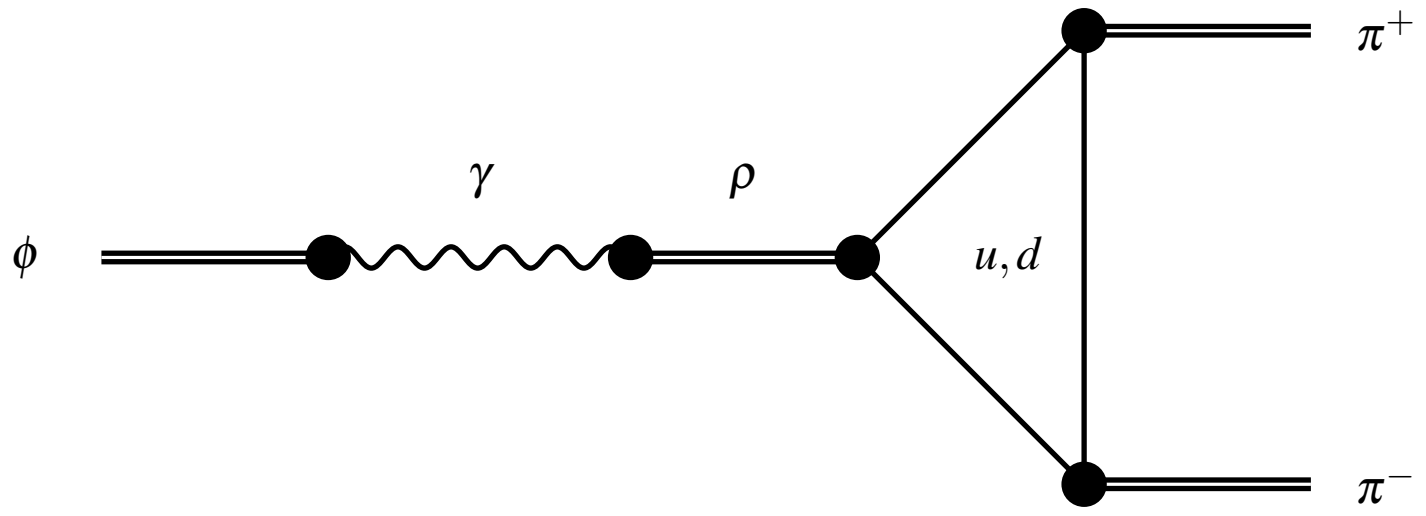%

Consider a channel with a direct $\phi - \rho$ transition, which can occur due to electromagnetic interaction with photon exchange and a non-zero difference in the masses of neutral and charged kaons.
The diagram with electromagnetic mixing $\phi - \rho$ is shown in Fig. \ref{phigamma}. The amplitude of this contribution has the form
\begin{eqnarray}
\label{phigamma}
    \mathcal{M}_{\gamma} = - \frac{\sqrt{2} 4 \pi \alpha_{em}}{3 g_\phi}
    p^2 BW_{\rho}(p)
    e_{\mu}^*(p) (p_{\pi^+}-p_{\pi^-})^\mu.
\end{eqnarray}

The decay branching fraction obtained with the amplitude of photon exchange in the $\phi - \rho$ transition is
\begin{eqnarray}
    Br(\phi\to \pi^+ \pi^-)_{\gamma} = 22.4 \times 10^{-5}.
\end{eqnarray}

Next we consider the direct $\phi - \rho$ transition channel that proceeds due to the contributions of the meson loop with neutral and charged kaons. It should be noted that the difference in the masses of $M_{K^\pm}$ and $M_{K^0}$ is primarily determined by the mass difference of $m_d-m_u$ \cite{Volkov:1986zb}. 
To calculate this contribution, we define the meson Lagrangian, which describes the decays of vector mesons into two kaons.
The amplitudes of vector meson-kaon transitions obtained from the Lagrangian (\ref{Lagrangian}) take the following form:

\begin{eqnarray}
    \mathcal{M}(V \to KK) & = & g_{VKK} e_\mu^*(p_\rho) (p_{K^i} - p_{K^j}),  
\end{eqnarray}
where $K^i=K^+(K^0)$, $K^j=K^-(\bar{K}^0)$. The decay constants are defined as $g_{\rho^0K^+K^-}=g_{\omega K^+K^-}=g_{\omega K^0 \bar{K}^0}=g_\rho/2$, $g_{\rho^0K^0\bar{K}^0}=-g_{\rho^0K^+K^-}$, $g_{\phi K^+K^-}=g_{\phi K^0 \bar{K}^0}=g_\phi/\sqrt{2}$. Then the meson Lagrangian describing the $V\to KK$ decays takes the form
\begin{eqnarray}
    \mathcal{L} = -i \frac{g_\rho}{2} \left[
    \left( K^- \partial_\mu K^+ - K^+ \partial_\mu K^- \right) \left( \rho^0 + \omega \right)^\mu  
    + \left( \bar{K}^0 \partial_\mu K^0 - K^0 \partial_\mu \bar{K}^0 \right) \left( - \rho^0 + \omega \right)^\mu 
    \right] \\ \nonumber
    - i \frac{g_\phi}{\sqrt{2}} \left[ \left( K^- \partial_\mu K^+ - K^+ \partial_\mu K^- \right) \phi^\mu +
    \left( \bar{K}^0 \partial_\mu K^0 - K^0 \partial_\mu \bar{K}^0 \right) \phi^\mu 
    \right].    
\end{eqnarray}

\begin{figure*}[t]
\begin{tikzpicture}
  \draw[draw={rgb,255:red,0;green,0;blue,0}, line width=1.00pt, double] (2.00, -4.00) -- (4.00, -4.00);
  \draw[draw={rgb,255:red,0;green,0;blue,0}, line width=1.00pt, double] (6.00, -4.00) -- (8.00, -4.00);
  \draw[draw={rgb,255:red,0;green,0;blue,0}, line width=1.00pt, double] (10.00, -4.00) -- (12.00, -4.00);
  \draw[draw={rgb,255:red,0;green,0;blue,0}, line width=1.00pt, double] (12.00, -4.00) -- (14.00, -4.00);
  \draw[fill=white!20, draw={rgb,255:red,0;green,0;blue,0}, line width=1.00pt, dashed, double] (5.00, -4.00) ellipse (1.00 and 1.00);
  \draw[fill=white!20, draw={rgb,255:red,0;green,0;blue,0}, line width=1.00pt, dashed, double] (12.00, -3.00) ellipse (0.95 and 0.95);
  \fill (4.00, -4.00) circle (0.15);
  \fill (6.00, -4.00) circle (0.15);
  \fill (12.00, -4.00) circle (0.15);
  \node at (1.50, -4.00) {$\phi$};
  \node at (8.50, -4.00) {$\rho$};
  \node at (9.50, -4.00) {$\phi$};
  \node at (14.50, -4.00) {$\rho$};
\end{tikzpicture}
  \caption{Diagrams describing the contribution of meson loops to the $\phi-\rho$ transition. The dashed line corresponds to neutral or charged kaons.}
  \label{kaonloop}
\end{figure*}
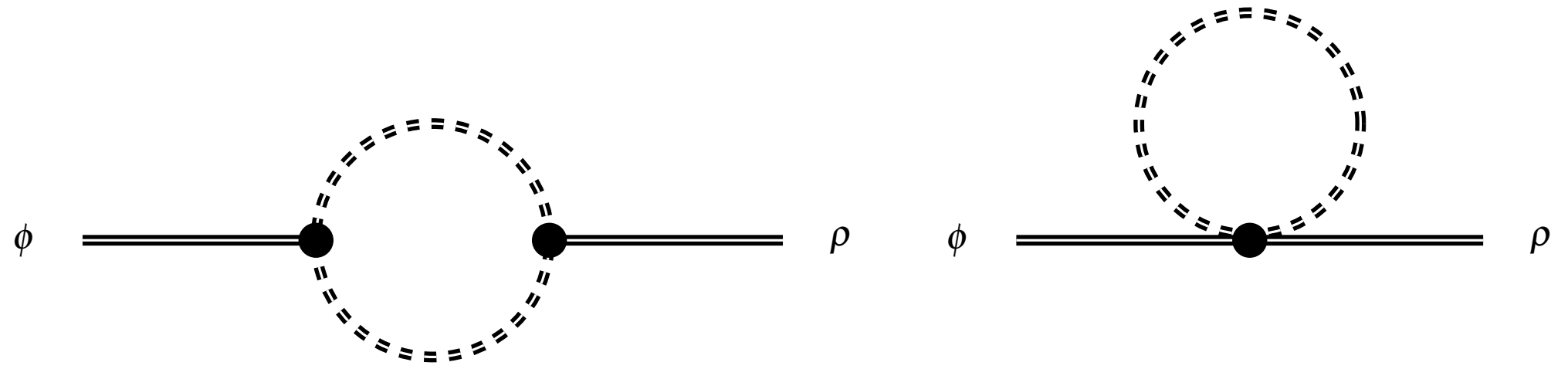%

A diagram with kaon loops of mixing $\phi - \rho$ is shown in Fig. \ref{kaonloop}.
The decay amplitude $\phi \to \pi^+ \pi^-$ with the kaon loop transition  $\phi - \rho$ has the form
\begin{eqnarray}
\label{phirho}
    \mathcal{M}_{\phi\rho} = 
    \frac{g^2_\rho g_\phi}{6\sqrt{2}}     
    \left[ I_{K^\pm} - I_{K^0} \right] BW_{\rho}(p)
    e_{\mu}^*(p) (p_{\pi^+}-p_{\pi^-})^\mu,
\end{eqnarray}
where $I_K^{\pm}$ is the kaon loop integral. This integral was obtained in Ref.~\cite{Volkov:2021fmi}:
    \begin{eqnarray}
    \label{Kloop}
		I_{K^\pm} & = & \frac{1}{16\pi^2} \left[2\Lambda_M^2 + \left(p^2 - 6M_{K}^2\right)\ln\left(1 + \frac{\Lambda_M^2}{M_{K}^2}\right) + 2\left(2\Lambda_M^2 + p^2 - 4M_{K}^2\right) D\left(\Lambda_M^2\right) \arctan\frac{1}{D\left(\Lambda_M^2\right)} \right.\nonumber\\
        && \left. + 2q^2 D(0)^3\arctan\frac{1}{D(0)}\right] - \frac{3}{8\pi^2} \left[\Lambda_M^2 - M_{K}^2 \ln\left(\frac{\Lambda_M^2}{M_{K}^2} + 1\right)\right],        
	\end{eqnarray}
where $D\left(\Lambda_M^2\right)=\sqrt{4 \left( \Lambda_M^2 + M_{K}^2\right)/p^2-1}$. The meson loop cutoff parameter is $\Lambda_M = 750$~MeV. Taking into account meson loops with this cutoff value led to reasonable corrections in the description of vector meson decays into a lepton pair \cite{Volkov:2021fmi}. The integral $I_{K^{0}}$ is obtained from $I_{K^{\pm}}$ by replacing the mass of a charged kaon with that of a neutral one.

For the contribution of the channel from the kaon loop of the $\phi - \rho$ transition to the decay branching fraction of the $\phi \to \pi^+ \pi^-$ decay, we obtain
\begin{eqnarray}
    Br(\phi\to \pi^+ \pi^-)_{\phi\rho} = 38.3 \times 10^{-5}.
\end{eqnarray}

Now we present the partial width of the second channel with the kaon loop of the $\phi$--$\rho$ transition together with electromagnetic mixing. There is negative interference between these diagrams, and the joint contribution takes the value
\begin{eqnarray}
    Br(\phi\to \pi^+ \pi^-)_{\phi\rho + \gamma} = 8.4 \times 10^{-5}.
\end{eqnarray}

When considering all contributions of the mixing amplitude $\phi$--$\omega$ (\ref{phiomega}) together with the transitions $\phi$--$\rho$ (\ref{phigamma}) and (\ref{phirho}), the contribution of the first channel is totally canceled by the interference of these channels. As a result, for the partial decay width $\phi \to \pi^+ \pi^-$, taking into account all the channels considered above, we obtain a theoretical estimate
\begin{eqnarray}
    Br(\phi\to \pi^+ \pi^-) = (8.4 \pm 1.26) \times 10^{-5}.
\end{eqnarray}

The uncertainty in the model's calculations is estimated at 15\%. The primary source is the partial conservation of axial current (PCAC). There are several other sources of uncertainty such as experimental errors in meson parameters. The 15\% estimate was obtained by comparing many previous NJL calculations with experimental data and takes into account all known sources of uncertainty including PCAC.

The obtained theoretical result is in satisfactory agreement with the experimental value \cite{ParticleDataGroup:2026mpi}
\begin{eqnarray}
    Br(\phi \to \pi^+ \pi^-)_{exp} = (9.5 \pm 1.9) \times 10^{-5}.
\end{eqnarray}

\section{Conclusion}
In conclusion, let us provide a brief analysis of the obtained results. The process $\phi \to \pi^+ \pi^-$ examined in the present work proceeds with the violation of both the OZI rule and isospin invariance. Two channels can be identified here. The experimentally known mixing of the $\phi-\omega$ states followed by the transitions into the $\rho$ meson is taken into account in the first channel.  
In the second case, the $\phi$ meson transits into a $\rho$ meson via a kaon loop, as well as through a photon, within the vector dominance model. The first channel yields a result of $2.8 \times 10^{-5}$, which is much lower than the experimental value. The second channel yields a result of $8.4 \times 10^{-5}$ that is in satisfactory agreement with the experimental data. The contribution of the first channel is almost entirely canceled out by destructive interference with the second channel. As a result, the final answer is $8.4 \times 10^{-5}$, which is in satisfactory agreement with the experiment.

It should be noted that while the second channel is described quite consistently in the framework of our model, the first channel relies on the experimentally established mixing of the $\phi$ and $\omega$ mesons, the nature of which is not exactly known. In particular, the exchange of three gluons can make a significant contribution to this mixing, in a manner analogous to what occurs in the decay $J/\psi \to \pi\pi$~\cite{BaldiniFerroli:2016mbs} (see also the references therein). In addition, the exchange of a vector tetraquark state can make a significant contribution. A similar situation occurred in the case of $\eta$-meson mixing. It was described using the 't Hooft interaction~\cite{tHooft:1976rip,Volkov:2005kw} or glueball exchange~\cite{Volkov:1986zb}.

The process $\phi \to \pi^+\pi^-$ was also considered in \cite{Achasov:1990at} using meson loops. This process was also examined within the framework of a chiral unitary approach in \cite{Oller:1999ag} taking into account an intermediate photon, a two-step $\phi-\omega-\rho$ transition, as well as $\phi-\rho$ mixing. Furthermore, this process was examined within the framework of chiral perturbation theory in~\cite{Kucukarslan:2006wk}, where the $\phi-\omega$ mixing parameter was determined using known experimental data.

\subsection*{Acknowledgments}
The authors are grateful to Prof. A. B. Arbuzov for useful discussions.


\begin{thebibliography}{99}
\bibitem{ParticleDataGroup:2026mpi}
F.~Takahashi \textit{et al.} [Particle Data Group],
Int. J. Mod. Phys. A \textbf{41} (2026) no.22, 2630011

\bibitem{Klingl:1996by}
F.~Klingl, N.~Kaiser and W.~Weise,
Z. Phys. A \textbf{356} (1996) no.2, 193-206

\bibitem{Benayoun:2007cu}
M.~Benayoun, P.~David, L.~DelBuono, O.~Leitner and H.~B.~O'Connell,
Eur. Phys. J. C \textbf{55} (2008), 199-236

\bibitem{Danilkin:2014cra}
I.~V.~Danilkin, C.~Fern{\'a}ndez-Ram{\'\i}rez, P.~Guo, V.~Mathieu, D.~Schott, M.~Shi and A.~P.~Szczepaniak,
Phys. Rev. D \textbf{91}, no.9, 094029 (2015)

\bibitem{Volkov:2020jor}
M.~K.~Volkov, A.~A.~Pivovarov and K.~Nurlan,
Mod. Phys. Lett. A \textbf{35} (2020) no.24, 2050200

\bibitem{Volkov:2021fmi}
M.~K.~Volkov, A.~A.~Osipov, A.~A.~Pivovarov and K.~Nurlan,
Phys. Rev. D \textbf{104}, no.3, 034021 (2021)

\bibitem{Volkov:1986zb} 
  M.~K.~Volkov,
  Sov.\ J.\ Part.\ Nucl.\  {\bf 17}, 186 (1986)
  [Fiz.\ Elem.\ Chast.\ Atom.\ Yadra {\bf 17}, 433 (1986)]

\bibitem{Achasov:2000pu}
M.~N.~Achasov, K.~I.~Beloborodov, A.~V.~Berdyugin, A.~V.~Bozhenok, A.~D.~Bukin, D.~A.~Bukin, S.~V.~Burdin, T.~V.~Dimova, V.~P.~Druzhinin and M.~S.~Dubrovin, \textit{et al.}
Phys. Lett. B \textbf{474} (2000), 188-193

\bibitem{CMD-3:2023alj}
F.~V.~Ignatov \textit{et al.} [CMD-3],
Phys. Rev. D \textbf{109} (2024) no.11, 112002

\bibitem{Nambu:1961tp}
Y.~Nambu and G.~Jona-Lasinio,
Phys. Rev. \textbf{122} (1961), 345-358

\bibitem{Eguchi:1976iz}
T.~Eguchi,
Phys. Rev. D \textbf{14} (1976), 2755

\bibitem{Ebert:1982pk}
D.~Ebert and M.~K.~Volkov,
Z. Phys. C \textbf{16} (1983), 205

\bibitem{Volkov:1984kq}
M.~K.~Volkov,
Annals Phys. \textbf{157}, 282-303 (1984)

\bibitem{Ebert:1985kz} 
  D.~Ebert and H.~Reinhardt,
  Nucl.\ Phys.\ B {\bf 271}, 188 (1986).

\bibitem{Vogl:1991qt} 
  U.~Vogl and W.~Weise,
  Prog.\ Part.\ Nucl.\ Phys.\  {\bf 27}, 195 (1991).

\bibitem{Klevansky:1992qe} 
  S.~P.~Klevansky,
  Rev.\ Mod.\ Phys.\  {\bf 64}, 649 (1992).

\bibitem{Volkov:1993jw}
  M.~K.~Volkov,
  Phys.\ Part.\ Nucl.\  {\bf 24} (1993) 35. 

\bibitem{Ebert:1994mf} 
  D.~Ebert, H.~Reinhardt and M.~K.~Volkov,
  Prog.\ Part.\ Nucl.\ Phys.\  {\bf 33}, 1 (1994).

\bibitem{Hatsuda:1994pi}
T.~Hatsuda and T.~Kunihiro,
Phys. Rept. \textbf{247} (1994), 221-367

\bibitem{Buballa:2003qv}
M.~Buballa,
Phys. Rept. \textbf{407}, 205-376 (2005)

\bibitem{Volkov:2005kw} 
  M.~K.~Volkov and A.~E.~Radzhabov,
  Phys.\ Usp.\  {\bf 49}, 551 (2006).

\bibitem{Volkov:2019awd}
M.~K.~Volkov and A.~A.~Pivovarov,
Phys. Part. Nucl. Lett. \textbf{16}, no.6, 565-568 (2019)

\bibitem{Ambrosino:2009sc}
F.~Ambrosino, A.~Antonelli, M.~Antonelli, F.~Archilli, P.~Beltrame, G.~Bencivenni, S.~Bertolucci, C.~Bini, C.~Bloise and S.~Bocchetta, \textit{et al.}
JHEP \textbf{07} (2009), 105

\bibitem{BaldiniFerroli:2016mbs}
R.~Baldini Ferroli, A.~Mangoni and S.~Pacetti,
Phys. Rev. C \textbf{98} (2018) no.4, 045210

\bibitem{tHooft:1976rip}
G.~'t Hooft,
Phys. Rev. Lett. \textbf{37}, 8-11 (1976)

\bibitem{Achasov:1990at}
N.~N.~Achasov and A.~A.~Kozhevnikov,
Int. J. Mod. Phys. A \textbf{7}, 4825-4854 (1992)

\bibitem{Oller:1999ag}
J.~A.~Oller, E.~Oset and J.~R.~Pelaez,
Phys. Rev. D \textbf{62}, 114017 (2000)

\bibitem{Kucukarslan:2006wk}
A.~Kucukarslan and U.~G.~Meissner,
Mod. Phys. Lett. A \textbf{21} (2006), 1423-1430
\end{thebibliography}
\end{document}